\documentstyle[12pt]{article}
\def\simlt{\mathrel{\lower2.5pt\vbox{\lineskip=0pt\baselineskip=0pt
           \hbox{$<$}\hbox{$\sim$}}}}
\def\simgt{\mathrel{\lower2.5pt\vbox{\lineskip=0pt\baselineskip=0pt
           \hbox{$>$}\hbox{$\sim$}}}}
\def\unity{{\hbox{1\kern-.8mm l}}}

\def\al{\alpha}
\def\be{\beta}
\def\ga{\gamma}
\def\ze{\zeta}

\def\to{\rightarrow}

\def\kev{\; {\rm KeV}}
\newcommand{\nue}{$\nu_e$}
\def\anue {\tilde{\nu}_e}
\def\anu {\tilde{\nu}}
\def\anuuno {\tilde{\nu}_1}
\def\anumu {\tilde{\nu}_\mu}
\def\anudue {\tilde{\nu}_2}
\def \epr {E^\prime}

\def \plb {Phys. Lett. B }
\def \npb {Nucl. Phys. B }
\def \prd {Phys. Rev. D }
\def \nue{\nu_e}
\def \numu{\nu_\mu}

\def \nuuno{\nu_1}
\def \nudue{\nu_2}

\newcommand{\EQ}{\begin{equation}}
\newcommand{\EN}{\end{equation}}

\newcommand{\bea}{\begin{eqnarray}}
\newcommand{\eea}{\end{eqnarray}}
\newcommand{\bean}{\begin{eqnarray*}}
\newcommand{\eean}{\end{eqnarray*}}

\begin{document}
\renewcommand{\thefootnote}{\fnsymbol{footnote}}
\setcounter{page}{1}
\begin{titlepage}
\begin{flushright}
\large
INFN-FE 04-94 \\
May 1994
\end{flushright}
\vspace{3.cm}
\begin{center}
{\large MAJORON  DECAY IN MATTER}\\
\vspace{0.5cm}
{\large Zurab G. Berezhiani${}^{(a,b)}$\footnote{E-mail:
31801::berezhiani, berezhiani@fe.infn.it } and
Anna Rossi${}^{(a)}$\footnote
{E-mail: 31801::rossi, rossi@fe.infn.it }}

\vspace{0.4cm}

{\em ${}^{(a)}$ INFN sezione di Ferrara, I-44100
Ferrara, Italy

\vspace{0.2cm}

${}^{(b)}$ Institute of Physics, Georgian Academy
of Sciences, 380077 Tbilisi, Georgia }

\vspace{0.3cm}

\end{center}
\vspace{1.0cm}
\begin{abstract}
It is well known that the matter  can significantly alter the
picture of neutrino oscillation \cite{W} or neutrino decay \cite{BV}.
Here we show that the presence of dense matter induces also the decay of
{\it massless} majoron, a Goldstone boson
associated with the spontaneous lepton number violation, into a couple of
neutrinos with the same (or in some cases also opposite) helicities.
We calculate the rates of such matter induced majoron decays in various
cases, depending on the neutrino type and the chemical content of the medium,
and analyse their  properties.

\end{abstract}
\end{titlepage}
\renewcommand{\thefootnote}{\arabic{footnote}}
\setcounter{footnote}{0}
\newpage

\vspace{0.4cm}

{\bf 1.} An attractive approach to the neutrino mass problem
is based on the concept
of the spontaneous violation of the global $B-L$ symmetry. As a result,
neutrinos acquire non-zero Majorana masses and the massless Nambu-Goldstone
appears in the particle spectrum, called the majoron \cite{CMP,GR}.
It could have interesting implications for particle physics and cosmology.
Since $B-L$ is free of axial anomalies, the majoron is strictly massless
and, therefore, is stable.

As far as $B-L$ is a global symmetry, there can be terms which break it
explicitly \cite{LMR}.
In particular, such terms could appear due to the non-perturbative quantum
gravity effects, in the form of higher order operators cutoff by the
Planck scale \cite{ABMS}.
As a consequence, the majoron $\chi$ gets non-zero mass, i.e. becomes a
pseudo-Goldstone boson, and then it could be unstable.
Moreover, in order to make such a massive majoron cosmologically safe,
its decay into neutrinos should be necessarily invoked \cite{ABMS}.
(For the case of the flavon models see also \cite{dario}.)
However, the appearance of the majoron mass is an artefact of some
theoretical speculations\footnote{In the context of certain more complex
models, the majoron can acquire non-zero mass (without any Planck scale
effect) at the quantum level due to axial anomalies. For example,
in the context of the models \cite{LPY}, majoron is simultaneously an
axion, in which case its decay into neutrinos could be possible.},
and it does not follow from any basic principle.
Once again, in the framework of the simplest model \cite{CMP} the
majoron is massless and, therefore, it is stable in vacuum.

In this letter we show that even in the case of strictly massless
majoron, its decay into a couple of neutrinos (or antineutrinos)
$\chi\to 2\nu$ (or $\chi\to 2\anu$), becomes possible in the presence
of enough dense matter background.
This phenomenon is analogous to the matter induced decay (MID) of
neutrino \cite{BV}. Coherent interactions with the medium lead
to an energy splitting between neutrino and antineutrino states,
providing available phase space for the emission of the majoron through
the decays $\nu\to \anu+\chi$ or $\anu\to \nu+\chi$, depending on which of
$\nu$ and $\anu$ states becomes "heavier" in the matter background.
The neutrino MID and its astrophysical implications were subsequently
studied in details in a number of papers \cite{BS,ChS,GKL,BMR}.
However, the fact that the majoron in itself is unstable in a medium and
can also undergo the MID was missed in these papers.\footnote{
In fact, such a possibility was first remarked by Rothstein, Babu and
Seckel in ref. \cite{ABMS}.}

Here we study in details the dynamics of the majoron matter induced
decays in various physical situations. The properties of the neutrino field
propagating in a matter background are given in the appendix.

\vspace{0.4cm}

{\bf 2.} Let us consider, for the sake of
simplicity, the case of one Majorana neutrino $\Psi=\nu+\anu$. The
neutrino state is identified with the left-handed spinor $\nu\equiv\nu_L$,
and the antineutrino state $\tilde{\nu}$
is defined as $\tilde{\nu}=\nu_{R}\equiv  C\bar{\nu}^T_L$.

The effect of the coherent neutrino scattering off matter constituents
can be described in terms of the matter potential $V$.
More specifically, we consider a neutrino propagating in the rest frame
of the unpolarized, constant density matter
(which is the lab frame in realistic situations).
In the framework of electroweak standard interactions, the matter potential
is given by \cite{W}:
\EQ
V=\frac{\sqrt{2}G_F}{m_n}\rho Y \label{poten}
\EN
where $G_F$ is the Fermi constant, $m_n$ is the nucleon mass,
$\rho$ is the matter density and $Y$ is a number which depends on the
neutrino type and the matter chemical content (namely, $Y=
Y_e-\frac{1}{2}Y_n$ for the $\nue$ state and $Y=-\frac{1}{2}Y_n$ for
$\nu_{\mu,\tau}$, where $Y_e, Y_n$ are the electron and
neutron number densities per nucleon: $Y_n=1-Y_e$ for electrically neutral
matter). The potential of the antineutrino state $\anu$ is $-V$.

The neutrino propagation in matter becomes equivalent to the
propagation in the presence of an external vector field
$V_\mu=(V,0)$ and can be described by the Dirac equation
\EQ
\label{eqofm1}
( i \partial_\mu \ga^\mu - m + V \ga^0\ga^5) \Psi = 0
\EN
Notice, that only the axial neutrino current interacts with the external
"matter" field $V_\mu$. Indeed, the the two chiral states in
$\Psi=\nu_L+\anu_R$ (which we have interpreted as neutrino and antineutrino)
have the opposite "charges" with respect to $V_\mu$.
As a result, the energy-momentum  dispersion relations are different for the
positive and negative helicity states (which in the case of the
ultra-relativistic neutrinos are essentially $\anu$ and $\nu$):
\bea
E_{+} & =  & \sqrt{(|{\bf p}|-V)^2 + m^2}
\label{enplus1}
\\
E_{-} & =  & \sqrt{(|{\bf p}|+V)^2 + m^2}
\label{enminus1}
\eea
This modification of the dispersion relations (see e.g. refs. \cite{NR})
as compared to the case of the propagation in vacuum, allows the 'heavier'
of the two helicity states to decay into the 'lighter' one
with emission of the massless boson.\footnote{
The eqs. (\ref{enplus1}), (\ref{enminus1}) can be rewritten as
$E^2_{\pm}=|{\bf p}|^2 +(m^{\pm}_{eff})^2$,
where $(m^{\pm}_{eff})^2=m^2\mp 2V|{\bf p}|+V^2$ are
the effective (momentum dependent)
'mass$^2$' terms for the $\pm 1$ helicity states.
The eventual negative  $m^2_{eff}$ (when $m^2<2V|{\bf p}|$)
is not in conflict with the causality principle, since the velocity is
not proportional to the momentum anymore and it
does not exceed the speed of light \cite{JP}.}
The natural candidate is the
majoron, with the following coupling to neutrinos:
\EQ
{\cal L}_\chi  =  i\,\frac{g}{2}\,\chi\bar{\Psi}\ga^5\Psi =
 i\,\frac{g}{2}\,\chi(\bar{\nu}_L\nu_R-\bar{\nu}_R\nu_L)
\label{nuchi}
\EN
For definiteness, let us consider the case $V>0$ (which is e.g. the
case of $\nu_e$ propagating in normal matter: $Y_n<0.66$).
In this situation the phase space is nonzero for the transition
$\nu\to\anu+\chi$. The corresponding width is \cite{GKL,BMR}:
\EQ
\Gamma=\frac{g^2 V}{8\pi}\left[\frac{2\xi+1}{\xi+1} -2\xi
\ln \left(\frac{\xi+1}{\xi}\right)\right]\, , \,\,\,\,\,
\xi=\frac{m^2}{4V E}
  \label{nugamma}
\EN
where $E$ is the energy of the initial neutrino. Clearly, when
$V<0$, as in the case of $\nu_{\mu,\tau}$ or in the case of $\nu_e$ in
strongly neutronized medium ($Y_n>0.66$), the $\anu\to \nu+\chi$ decay
occurs. The neutrino MID exhibites
unusual energy dependence of neutrino lifetime in lab frame:
$\tau_{MID}\propto E^{-2}$ up to $E\sim m^2/V$ and then, with further
increasing $E$, slowly decreases to a constant value $\Gamma=g^2 V/8\pi$.
This is just the opposite to the case of usual decay in vacuum
where the slow particles decay faster: $\tau_{VAC}\propto E$.
Thus, In the limit of very light neutrino, $m^2\ll E V$, the decay width is
determined essentially by the matter potential \cite{BV,BS}.

On the other hand, if the matter is sufficiently dense, the majoron decay
$\chi\to 2\anu$ is also allowed.
Indeed, the matrix element is the same as that of the neutrino MID, and the
phase space considerations are similar.
 Let us calculate now the MID width of the majoron with four-momentum
$k=(\omega,{\bf k})$, propagating in the medium with  constant density.
We use the explicit solutions of the Dirac equation (\ref{eqofm1})
which are given in Appendix. $p=(\epr,{\bf p})$ and $q=(E,{\bf q})$ are
the four-momenta of the emitted $\anu$'s. We have:
\EQ
d\Gamma=\frac{1}{2}\frac{d\Phi}{2\omega}|M|^2 \label{digamma}
\EN
where the factor 1/2
is due to the identical final states. For the squared matrix element, we
obtain, neglecting the small terms ${\cal O}(m^2/E^2)$ and ${\cal O}(V/E)$:
\EQ
|M|^2=g^2 |C(q,p)|^2 |\al({\bf q})^+ \be({\bf p})|^2 \cong
2\,g^2\,|{\bf p}|\,|{\bf q}|\,(1-\cos\theta)  \label{emme}
\EN
where $\theta$ is the angle between ${\bf p}$ and ${\bf q}$ momenta and
\bea
C(q,p) &\!\!=&\!\! \sqrt{(E-|{\bf q}|+V)(\epr-|{\bf p}|+V)}
 \nonumber \\
& & -\sqrt{(E+|{\bf q}|-V)(\epr+|{\bf p}|-V)}
\simeq -2\sqrt{|{\bf q}||{\bf p}|}
\eea
The differential MID rate is easily obtained by carrying out the trivial
integrations in the phase space. Due to the four-momentum conservation
the r.h.s. of eq. (\ref{emme}) becomes
\EQ
g^2\left[\left(|{\bf p}|+|{\bf q}|\right)^2-\omega^2\right] \simeq
 g^2 4\omega V\left[1-x \omega
\left(\frac{1}{\omega-E} +\frac{1}{E}\right)\right]
\label{emme1}
\EN
so that we have
\EQ
\frac{d\Gamma}{d E}=\frac{g^2V}{8\pi\, \omega}
\left[1-x \omega\,\left(
\frac{1}{\omega- E} +\frac{1}{E}\right)\right]\,,
\,\,\,\,x=\frac{m^2}{4V\omega }
\label{dege}
\EN
The physical range of the emitted $\anu$ energies   is
\EQ
\frac{\omega}{2}\left(1-\sqrt{1-4x}\right) \leq E,\epr\leq
\frac{\omega}{2}\left(1+\sqrt{1-4x}\right)
\EN
Thus, that the MID $\chi\to 2\anu$ can occur if the majoron energy is above
the threshold energy $\omega_{th}=m^2/V$. The total width has the form:
\EQ
\Gamma  = \frac{g^2V}{8 \pi} \,\left[
\sqrt{1-4x}+2x\ln \left(\frac{
1-\sqrt{1-4x}}{1+\sqrt{1-4x}}\right)\right] \label{rate}
\EN

In the limit $x\ll 1$ when the matter potential term dominates over the
neutrino mass ( i.e. $\omega V \gg m^2$), the energy distribution of the
emitted (anti)neutrinos is flat within the energy range
$\frac{m^2}{4V}\leq E\leq \omega-\frac{m^2}{4V}$:
\footnote{We remind that we have neglected ${\cal O}(m^2/E^2)$ terms in
$|M|^2$. This is safe since in
the  realistic physical situations, when $m\gg V$,
the final states are always relativistic. If $m\sim V$, one of the final
states can be non-relativistic, but the corresponding phase space portion
is very small ($\sim V$) and it induces only $O(V^2/E)$ corrections to the
decay width (\ref{rate}).}
\EQ
\frac{d\Gamma}{d E}=\frac{g^2V}{8\pi\, \omega} \label{major}
\EN
and the total width is independent of the majoron energy:
\EQ
\Gamma  =  \frac{g^2V}{8 \pi} \label{common}
\EN
This expression coincides with the one for the neutrino MID
$\nu\to \anu +\chi$ in the limit of very light neutrinos.
For clearness, we remind that the $\chi\to 2\anu$ occurs if $V>0$.
When $V<0$, the $\chi\to 2\nu$ decay takes place.

\vspace{0.4cm}

{\bf 3.} Until now we have considered
the case of only one Majorana neutrino $\nu$ with matter potential $V$.
Our considerations can be applied to the case of ZKM  or
Dirac neutrinos as well. Let us consider, for example,
the case of ZKM neutrino $\Psi=\nue+ \anumu$, resulting from the
spontaneous violation of $U(1)_e\otimes U(1)_\mu$ global symmetry down
to $U(1)_{e-\mu}$.
Here $\nue=\nu_L$ is the electron neutrino (L.H.) state
with the lepton number $L_e=1$ and matter
potential $V_e$, and $\anumu=\nu_R=C\bar{\nu}^{T}_\mu$ is a R.H.
state of the muon antineutrino
with the matter potential $-V_\mu$.
The mass term $m\bar{\Psi}\Psi=m\bar{\nu}_e\anumu +$h.c. and
the neutrino coupling with the majoron
$i\frac{g}{2}\chi\bar{\nu}_e\anumu+$h.c.
conserve the lepton number combination $L_{-}=L_{e} -L_\mu$.
In the case of normal matter we have $V_e>|V_\mu|$.
Therefore, the neutrino MID can
proceed  through both the helicity flipping (h.f.) decay
 $\nue\to\anumu+\chi$ and the
helicity conserving (h.c.) one $\nue\to\numu+\chi$.
For the matrix elements of these decay modes we have, at the leading order,
the following:
\EQ
|M|_{hf}
^2\approx 2g^2 |{\bf p}||{\bf q}| (1-\cos\theta) \label{eqhf}
\EN
\EQ
|M|_{hc}
^2\approx \frac{1}{2}g^2 m^2\left(\frac{|{\bf p}|}{|{\bf q}|}
+\frac{|{\bf q}|}{|{\bf p}|} -2\right) (1+\cos\theta) \label{eqh}
\EN
where $p=(\epr,{\bf p})$ and $q=(E,{\bf q})$ are the four-momenta of the
initial and final fermions respectively and $\theta$ is the angle between
${\bf p}$ and ${\bf q}$. Then the differential decay rates are:
\EQ
\frac{d\Gamma}{d\epr}\left(\nue\to\anumu\chi\right)  =
\frac{g^2m^2}{16\pi E^2}\left(1-\frac{\epr}{E}\right)
\left(\frac{4VE}{m^2} +1 -\frac{E}{\epr}\right) \label{hfdif}
\EN
\EQ
\frac{d\Gamma}{d\epr}\left(\nue\to\numu\chi\right) =
\frac{g^2m^2}{16\pi E^{2}}
\left(\frac{\epr}{E}+\frac{E}{\epr} -2\right) \label{hcdif}
\EN
and for the physical range of the final fermion energy we have
\EQ
\frac{m^2}{m^2+4V_{\pm}E}E\leq \epr \leq E
\EN
where $V_{+}=\frac{1}{2}(V_e + V_\mu)$
(for the helicity flipping case) and
$V_{-}=\frac{1}{2}(V_e - V_\mu)$ (for  the helicity conserving one).
The partial widthes are given by:
\EQ
\Gamma\left(\nue\to\anumu\chi\right)=\frac{g^2 V_+}{8\pi}
\left[\frac{2\xi+1}{\xi+1} -2\xi
\ln \left(\frac{\xi+1}{\xi}\right)\right]\,,\,\,\,
\xi=\frac{m^2}{4V_+ E}  \label{nuhf}
\EN
\EQ
\Gamma\left(\nue\to\numu\chi\right)=
\frac{g^2 V_-}{8\pi}\ze\left[\ln \left(
\frac {\ze+1}{\ze}\right)-\frac{\ze+3/2}{(\ze+1)^2}\right] \,,\,\,\,
\ze=\frac{m^2}{4V_- E}\label{nuhc}
\EN
Therefore, in the 'small mass'  limit ($\xi,\ze\ll 1$) the h.f.
decay is dominant, and its width depends only on the matter potential:
$\Gamma\left(\nue\to\anumu\chi\right)=g^2 V_{+}/8\pi$
(cfr. eq. (\ref{major})),
whereas the h.c. one vanishes as $m\to 0$.
In the opposite regime of 'large mass' ($\xi,\ze\gg 1$) the two widthes
have the same shape:
\EQ
\frac{\Gamma\left(\nue\to\anumu\chi\right)}
{V^3_+}
=\frac{\Gamma\left(\nue\to\numu\chi\right)}
{V^3_-}=\frac{2}{3}\frac{g^2 E^{2}}{\pi m^4}
\EN
Both of these are suppressed (by factors $\xi^3$ and $\zeta^3$ as compared to
the one of eq. (\ref{major})) and their values are comparable.

Let us consider the
majoron MID channels in the case of ZKM neutrino.
There is a direct correspondence with the neutrino MID matrix elements due to
the "crossing" property.
If  $V_e+V_\mu$ is positive, the majoron MID
$\chi\to \tilde{\nu}_e+\anumu$ is allowed.
The matrix element is still given by the formula (\ref{eqhf}) and the decay
rate is given by the formulas similar to (\ref{dege}) and (\ref{rate}):
\EQ
\Gamma\left(\chi\to\tilde{\nu}_e\anumu\right)
= \frac{g^2 V_+}{4\pi} \,\left[
\sqrt{1-4x}+2x\ln \left(\frac{
1-\sqrt{1-4x}}{1+\sqrt{1-4x}}\right)\right]
\,,\,\,\,x=\frac{m^2}{4V_+\omega} \label{ratezkm}
\EN
Since $V_e-V_\mu>0$, the majoron MID with the opposite helicity final
fermions is $\chi\to  \anue +\numu$.
Its matrix element is similar to the one of eq. (\ref{eqh})
and the differential and total rates are the folowing:
\EQ
\frac{d\Gamma}{dE}\left(\chi\to \anue \numu\right)
=\frac{g^2m^2}{16\pi \omega^2}
\left( \frac{\omega-E}{E}+\frac{E}{\omega-E} +2\right)
\EN
\EQ
\Gamma\left(\chi\to \anue \numu\right)
= \frac{g^2V_-}{2\pi} z
\ln\left(\frac{1+\sqrt{1-4z}}{1-\sqrt{1-4z}}\right)
\label{seconda}\,,\,\, \,\,
  z= \frac{m^2}{4 V_-\omega}
\EN
Thus, in the limit $m\to 0$ the h.c. MID rate
$\Gamma(\chi\to \anue \numu)$
is vanishing whereas the h.f. one is maximal:
$\Gamma(\chi\to \anue \anumu)= g^2V_{+}/4\pi$.


The Dirac case is recovered when the right handed state $\nu_R$ is sterile
i.e. $V_R=0$. In this case the two majoron MID modes $\chi\to \anue +
\tilde{\nu}_R$ and $\chi\to \anue+\nu_R$ can be directly compared
($V_{+}=V_{-}$).

The straightforward generalization for the case of the majoron matter
induced decays
$\chi\to\anuuno +\anudue$ and $\chi\to\anuuno +\nudue$, where $\nuuno$ and
$\nudue$ are two neutrinos
with different masses $m_1, m_2$
and different matter potentials $V_1, V_2$, leads to
\EQ
\Gamma\left(\chi\to\tilde{\nu}_1\tilde{\nu}_2\right)
= \frac{g^2V_+}
{4 \pi} F(x_1, x_2)\,\,,\,\,\,\, x_i  = \frac{m^2_i}{4V_+\omega}
\label{gamma}
\EN
\begin{displaymath}
F(x_1, x_2)  =
f(x_1, x_2)+
x_1 g_1(x_1,x_2)+x_2 g_2(x_1,x_2)\nonumber
%
\end{displaymath}
\EQ
\Gamma\left(\chi\to\anuuno\nudue\right)
=  \frac{g^2V_-}{4\pi}\sqrt{z_1z_2} G(z_1,z_2)\,\,,\,\,\,\,\,
 z_i  = \frac{m^2_i}{4V_- \omega}
\EN
\begin{displaymath}
G(z_1,z_2)  =
\left(2 -\frac{m_1}{m_2}-\frac{m_2}{m_1}\right) f(z_1,z_2)
-\frac{m_1}{m_2}g_1(z_1,z_2)- \frac{m_2}{m_1}g_2(z_1,z_2)
\end{displaymath}
where $V_{\pm}=\frac{1}{2}(V_1\pm V_2)$ and
\bea
& & f(a,b)=\sqrt{(1-a-b)^2-4 a b} \nonumber\\
& & g_{1,2}(a,b) = \ln\left(
\frac{1\pm (a-b)-f(a,b)}{1\pm (a-b)+f(a,b)}\right)
\eea

\vspace{0.4cm}

{\bf 4.}
In order to give some numerical insight on  such matter induced decays,
let us consider  the limit of small neutrino mass. Then for the
lifetimes of the neutrino and majoron MID, we have
\EQ
\tau_\nu=\tau_\chi\sim
(0.4~\mbox{s})\left(\frac{1~\mbox{g}/\mbox{cm}^3}{g^{2}\rho Y}\right)
\EN
For example,  for the typical supernova densities $\rho\sim 10^{14}
\mbox{g/cm}^3$, by comparing the above
equation\footnote{In fact, in the case of supernova
the "small mass" limit $\xi\ll 1$ is valid if $m< 10 \kev$.}
to the diffusive cooling time $\tau\sim 1$s, we see that these MID
effects can be relevant down to the values $g \sim 10^{-7}$, which
is well below any laboratory and astrophysical limits on
the $\nu-\chi$ coupling constants \cite{kim,kolb}.
However, considering the MID effects for the supernova, one has to take
into account that neutrinos are in equilibrium in the core and
comprise a certain amount of the total particle number density.
Hence they are contributing to the
matter potential $V$ in eq. (\ref {poten}),
as well as providing the Pauli blocking.
For example, by taking the case of electron neutrino $\nu=\nue$, we have
for $Y=Y_e -\frac{1}{2}Y_n +2Y_\nu=\frac{3}{2}
Y_l -\frac{1}{2}(1-Y_\nu)$, where $Y_l=Y_e+Y_\nu$ is the density of lepton
number trapped in the supernova core.
At the stage of diffusive cooling, due to the strong neutronization of
shocked matter, we have typically $Y<0$. Therefore the relevant MID's
are $\anue\to\nue+\chi$ and $\chi\to 2\nu_e$.
However, the formulas (\ref{nugamma}) and (\ref{rate})
cannot be applied directly in this case, since
one has to take into account the thermal bath of $\nue$'s with chemical
potential $\mu$ and temperature $T$. Therefore, considering either
$\anue\to\nue+\chi$ or $\chi\to 2\nu_e$
decays, one has to include the
factor $(1-f_\eta (E/T))$ in the phase space per each $\nue$ final state,
where $f_\eta$ is the Fermi-Dirac distribution function:
\EQ
f_\eta(\frac{E}{T})=\frac{1}{\exp(\frac{E-\mu}{T}) +1} \,\,, \,\,\,\,\,\,
\eta=\frac{\mu}{T}
\EN
Then for the MID differential rates we have
\EQ
\frac{d\Gamma}{d\epr}\left(\tilde{\nu}_e\to\nue\chi\right) \!=\!
\frac{g^2m^2}{16\pi E^{2}}\left(1-\frac{\epr}{E}\right)
\left(\frac{4VE}{m^2} +1 -\frac{E}{\epr}\right)
\left(1-f_\eta(\frac{\epr}{T})\right)
\EN
\EQ
\frac{d\Gamma}{d E}(\chi\to 2\nue)\!\!=\!\!
\frac{g^2V}{8\pi\, \omega}
\left[1-\frac{m^2}{4V}\,\left(
\frac{1}{E} +\frac{1}{\omega- E}\right)\right]
\left(1-f_\eta(\frac{E}{T})\right)\!
\left(1-f_\eta(\frac{\omega-E}{T})\right )
\label{dege1}
\EN
In both the above equations, the value for the
matter potential V refers to the 'heavier'
state in matter, i.e. $\anue$.

Once again, in the supernova core we can neglect the neutrino mass
-- $m^2\ll EV$.
Then for the MID rates we obtain
\bea
\Gamma\left(\tilde{\nu}_e\to\nue\chi\right) &\!\! =\!\! &
\frac{g^2V}{8\pi\,\tilde{E}}
\left\{
\frac{2}{T} \int^{E}_{0} \mbox{d} \epr \frac{\epr}{E}
f_\eta (\frac{\epr}{T}) - \tilde{E} \right. \nonumber \\
& & \left. -2\ln \left[f_\eta(\frac{E}{T})(\mbox{e}^{-\eta} +1)\right]
 \right\} \,\, ,\,\,\,\,\,\,\,\,
\tilde{E}\!=\!\frac{E}{T} \label{ferminu}
\eea
\EQ
\Gamma(\chi\to 2\nue)
=\frac{g^2V}{8\pi\,\tilde{\omega}}\frac{1}{\left[\exp(2\eta-\tilde{\omega})
-1\right]} \ln \left[\frac{1+\cosh\eta}{1+\cosh(\tilde{\omega}-\eta)}
\right]\,\,,\,\,\,\,\,\tilde{\omega}=\frac{\omega}{T}  \label{fermichi}
\EN
In the case of strongly degenerated neutrino thermal background
($\eta\gg 1$), the above equations imply for $\tilde{E}, \tilde{\omega}
<\eta$
\EQ
 \Gamma\left(\tilde{\nu}_e\to\nue\chi\right)  =
\frac{g^2V}{8\pi}\,\frac
{\mbox{e}^{\tilde{E}}-\tilde{E}-1}{\tilde{E}^2}\,\mbox{e}^{-\eta}
\EN
\EQ
\Gamma(\chi\to 2\nue) = \frac{g^2 V}{8\pi} \mbox{e}^{\tilde{\omega}-2\eta}
\EN
whereas for $\tilde{E},\tilde{\omega}> \eta$ we have
\EQ
 \Gamma\left(\tilde{\nu}_e\to\nue\chi\right)  =
\frac{g^2 V}{8\pi}\left(1-\frac{2\eta}{\tilde{E}}
\right)
\EN
\EQ
\Gamma(\chi\to 2\nue)=\frac{g^2 V}{8\pi}\left(1-\frac{2\eta}{\tilde{\omega}}
\right)
\EN
When $\eta \ll 1$,
and $E, \omega<T$, the neutrino and majoron MID are
$\Gamma\left(\tilde{\nu}_e\to\nue\chi\right) \!=\!g^2 V/16\pi$  and
$\Gamma(\chi\to 2\nue)=g^2 V/32\pi$.
Hence, the presence of the neutrino thermal bath brings just suppression
factors 2 and 4  for the neutrino and majoron MIDs respectively,
as compared to the eq. (\ref{common}). When $E, \omega>T$, there is no
suppression and both MID widthes are given by eq. (\ref{common}).

\vspace{0.4cm}

{\bf 5.} It is well known that the coherent interaction with matter removes
the energy degeneracy of different helicity neutrino states. Hence
the neutrino, even if it is stable in vacuum, can undergo the matter
induced decay into the antineutrino and the {\it massless} majoron.
However, the majoron in itself is unstable in matter and decays in a couple
of antineutrinos. The matrix element and phase space considerations
are similar for both the neutrino and majoron MIDs and
the corresponding widthes are comparable. In this paper the features
of the majoron MID have been studied for various neutrino types.
We have also calculated the rates for the neutrino and majoron
MIDs in the presence of thermal neutrino bath, as it occurs in the
supernova core.
The impact of the neutrino MID on the supernova dynamics and
$\nu$-signal was studied in refs. \cite{BS,ChS}, but the effects of the majoron
MID were not included. We have seen that these effects can be important
for $\nu-\chi$ coupling constants $g> 10^{-7}\div 10^{-6}$.
In the 'standard' model of the singlet majoron  \cite{CMP} such a 'large'
constant is rather improbable. However, there is
a variety of singlet majoron models \cite{BSV} which could naturally
provide $g$ in this range. The effects of the majoron MID for
the supernova will be studied elsewhere.

Last but not least, let us remark the following. In fact,
the majoron could receive considerable mass $M$,
e.g. due to the  'Planck scale' terms explicitly violating the lepton number.
Then the neutrino MID  $\nu\to\tilde{\nu}
+\chi$  could not occur. However, in the medium there is still available
phase space for the neutrino 3-body
decay $\nu\to 3 \tilde{\nu}$  which can be mediated by the
exchange of such a heavy 'majoron' -- or, in other words, due to the
lepton number violating 4-fermion interaction $\frac{1}{M^2}(\nu C\nu)^2$.
The properties of such a 3-body MID will be given elsewhere.

\vspace{1cm}

{\bf Acknowledgments.}

\vspace{0.5cm}

We thank Georg Raffelt and Dave Seckel for reading the manuscript and
useful comments. Discussions with Jim Cline, Gianni Fiorentini and
Semen Gershtein are also gratefully acknowledged.

\vspace{1.cm}

\section*{Appendix}
\setcounter{equation}{0}
\renewcommand{\theequation}{A.\arabic{equation}}

\vspace{0.5cm}

We look for the solutions of the Dirac equation
$(  p_\mu \ga^\mu - m  + V \ga^0\ga^5) u ({\bf p}) = 0 $
in the form of plane waves with negative frequency $u({\bf p})\exp(-ipx)$
and positive frequency $v({\bf p})\exp(+ipx)$.
Let us consider in some detail the former case, where we have:
\begin{displaymath}
\left(
\begin{array}{cc}
m  & E + {\bf p}\cdot{\bf \sigma} - V
\\
E - {\bf p}\cdot{\bf \sigma} + V  & m
\end{array}
\right)
\left(
\begin{array}{c}
u^1 ({\bf p}) \\ u^2 ({\bf p})
\end{array}
\right)
= 0
\end{displaymath}
Here the Weyl representation for the gamma matrices is adopted;
$\sigma^i$ are the usual $2\times 2$ hermitian Pauli matrices.
Since the helicity operator $\frac{{\bf p}\cdot \sigma}{|{\bf p}|}$ is still
a constant of motion, one can consider the solutions $u_{+}({\bf p})$ and
$u_{-}({\bf p})$ with different helicities.
The presence of the matter potential
determines a distinct energy eigenvalue for each helicity eigenstate,
in contrast with the usual energy degeneracy in vacuum.
The spinors  $u_{\pm}({\bf p})$ (normalized as
$u_{\pm}({\bf p})^+ u_{\pm}({\bf p}) = 2 E_{\pm}$) are the following:
\begin{displaymath}
u_{+}({\bf p})\!=\! \left(\begin{array}{c}
-\sqrt{E_{+}+|{\bf p}|-V}\, \alpha({\bf p}) \\
\sqrt{E_{+}-|{\bf p}|+V}\, \alpha({\bf p})
\end{array}\right)
\;\;  ,  \;\;
u_{-}({\bf p})\!=\! \left(\begin{array}{c}
-\sqrt{E_{-}-|{\bf p}|-V}\, \beta({\bf p}) \\
\sqrt{E_{-}+|{\bf p}|+V}\, \beta({\bf p})
\end{array}\right)
\end{displaymath}
where $\al({\bf p})$ and $\be({\bf p})$ are the $+1$ and $-1$
helicity normalized eigenspinors, respectively.
The matrices $\al({\bf p})\al({\bf p})^+$
and $\be({\bf p})\be({\bf p})^+$ are helicity projection operators:
\begin{displaymath}
\al({\bf p})\al({\bf p})^+=\frac{1}{2}
\left( \unity + \frac{{\bf p}\cdot{\bf \sigma}}{|{\bf p}|}\right)
\;\; , \;\;
\beta({\bf p}) \beta({\bf p})^+ =
\frac{1}{2}\left( \unity - \frac{{\bf p}\cdot{\bf \sigma}}{|{\bf p}|}\right)
\end{displaymath}

The normalized positive frequency solutions are
obtained in an analogous way:
\begin{displaymath}
v_{+}({\bf p})\!=\! \left(\begin{array}{c}
\sqrt{E_{+}-|{\bf p}|+V} \,\beta({\bf p}) \\
\sqrt{E_{+}+|{\bf p}|-V} \,\beta({\bf p})
\end{array}\right)
\;\;  ,  \;\;
v_{-}({\bf p})\!=\! \left(\begin{array}{c}
-\sqrt{E_{-}+|{\bf p}|+V} \,\alpha({\bf p}) \\
-\sqrt{E_{-}-|{\bf p}|-V} \,\alpha({\bf p})
\end{array}\right)
\end{displaymath}

The energy-momentum dispersion relations  for the positive and
negative helicity solutions are respectively given by
the eqs. (\ref{enplus1}) and (\ref{enminus1}).
The helicity eigenspinors can be chosen in such a way that
$\alpha({\bf p})$ and $\beta({\bf p})$ are related to each other
by charge conjugation:
\begin{displaymath}
i\sigma^2 \alpha({\bf p})^* = \beta({\bf p})  \;\; , \;\;
i\sigma^2 \beta({\bf p})^* = -\alpha({\bf p})  \;\; .
\label{alfbet}
\end{displaymath}
Summarizing, the (Majorana) solution of eq. (\ref{eqofm1}) has the form
\bean
\Psi = \Psi^c &  = &
{\bf \sum_p}\left(
u_{+}({\bf p})f_{+}({\bf p})e^{-ip_{+}x}
+u_{-}({\bf p})f_{-}({\bf p})e^{-ip_{-}x} \right.
\\
& & \phantom{\bf \sum_k} \left.
+v_{+}({\bf p}){f}^{\dagger}_{+}({\bf p})e^{+ip_{+}x}
+v_{-}({\bf p}){f}^{\dagger}_{-}({\bf p})e^{+ip_{-}x}\right)
\eean
where $p_{\pm}^{\mu}=(E_{\pm},{\bf p})$ and $f_{\pm}$ and
$f^{\dagger}_{\pm}$ are the annihilation and creation operators
(related to each other by charge conjugation)  for particle
with helicity $\pm 1$. Therefore,
this Majorana field describes the creation and annihilation of two particle
states only, one characterized by helicity $+1$ and energy $E_{+}$
(eq. (\ref{enplus1})), the other  characterized by helicity $-1$ and energy
$E_{-}$ (eq. (\ref{enminus1})).
The former is conventionally
referred as `antineutrino' and the latter as `neutrino'.


\end{document}